\documentclass[12pt]{article}
\usepackage[utf8]{inputenc}
\usepackage{graphicx}
\usepackage{xcolor}
\usepackage{natbib}
\usepackage[toc,page]{appendix}
\usepackage{geometry}
\usepackage{microtype}
\usepackage{setspace}
\usepackage{fullpage}
\usepackage{lineno}

\title{Fast and reliable confidence intervals for a variance component}
\author{Yiqiao Zhang$^\star$ \quad Karl Oskar Ekvall$^{\star, \dagger}$ \quad
Aaron J. Molstad$^{ \ddagger}$\\
{\normalsize $^\star$Department of Statistics, University of Florida} \\
{\normalsize $^\dagger$Division of Biostatistics, Institute of Environmental
Medicine, Karolinska Institutet}\\
{\normalsize $^\ddagger$School of Statistics, University of Minnesota}\\
{\tt \normalsize yiqiaozhang@ufl.edu \quad k.ekvall@ufl.edu \quad
amolstad@umn.edu}}
\date{\normalsize \today}
\date{\today} 

\usepackage{hyperref}
\usepackage{lipsum}

\graphicspath{{./figures/}}

\usepackage{amsmath, amssymb, amsthm}
\theoremstyle{plain}
\newtheorem{lemma}{Lemma}
\newtheorem{theorem}{Theorem}
\newtheorem{proposition}{Proposition}
\newtheorem{corollary}{Corollary}
\theoremstyle{definition}

\theoremstyle{remark}

\DeclareMathOperator{\diag}{diag}

\newcommand{\cov}{\mathrm{cov}}

\newcommand{\pr}{\mathrm{P}}

\newcommand{\tr}{\operatorname{tr}}
\newcommand{\tsp}{\mathrm{\scriptscriptstyle T}}
\newcommand{\rN}{\mathrm{N}}

\newcommand{\R}[1]{\mathbb{R}^{#1}}

\def\T{{ \mathrm{\scriptscriptstyle T} }}
\usepackage{cleveref}

\begin{document}
\maketitle

\onehalfspacing

\begin{abstract}
We show that confidence intervals in a variance component model, with
asymptotically correct uniform coverage probability, can be obtained by
inverting certain test-statistics based on the score for the restricted
likelihood. The results apply in settings where the variance
is near or at the boundary of the parameter set. Simulations indicate the
proposed test-statistics are approximately pivotal and lead to confidence
intervals with near-nominal coverage even in small samples. We illustrate
our methods' application in spatially-resolved transcriptomics where we
compute approximately 15,000 confidence intervals, used for gene ranking, in
less than 4 minutes. In the settings we consider, the proposed method is
between two and 28,000 times faster than popular alternatives, depending on
how many confidence intervals are computed.
\end{abstract}

\section{Introduction} \label{sec:Introduction}

Variance component models are widely used in applied statistics. Often,
researchers are interested in quantifying how much of the variability in the
data is due to a specific source, possibly after controlling for, or
conditioning on, other sources of variability. It is
common to assume that for non-stochastic $X \in \R{n\times p}$ and $Z \in
\R{n\times q}$, a vector of responses $Y \in \R{n}$ satisfies
\begin{equation} \label{eq:lmm}
    Y = X\beta + ZU + E,
\end{equation}
where $U\sim \rN_q(0, \sigma_g I_q)$ and, independently, $E \sim \rN_n(0,
\sigma^2_e I_n)$. Both $U$ and $E$ are unobservable, and the parameter is
$\theta = (\beta, \sigma^2_g, \sigma^2_e) \in \R{p}\times [0, \infty) \times (0,
\infty)$. Equation \eqref{eq:lmm} is a linear mixed model
with one random effect, or several independent random effects with the same
variance.

In modern genetics and genomics, the model is often written directly for the marginal distribution of $Y$:
\begin{equation}\label{eq:one_varcomp}
Y \sim \mathcal{N}_n(X\beta, \sigma_g^2 K + \sigma_e^2 I_n),
\end{equation}
where $K = ZZ^\tsp$;
see for example
\citet{purcell2002variance,kang2010variance,zhang2010mixed,lippert2011fast,
yang2011gcta,zhou2012genome,zhou2013polygenic,loh2015efficient,
weissbrod2016multikernel,moore2019linear}. Equations
\eqref{eq:lmm} and \eqref{eq:one_varcomp} give equivalent models since, for any positive
semi-definite $K \in \R{n\times n}$ of rank $q \leq n$, there exists a $Z \in
\R{n\times q}$ such that $K = ZZ^\tsp$. In statistical genetics, when elements
of $Y$ are phenotypes for $n$ individuals, $K$ can be a genetic relatedness
matrix quantifying similarities between individuals' common SNP genotypes
\citep{jiang2022unbiased}. Then, the quantity of inferential interest is the
heritability \citep{visscher2008heritability}
\begin{equation*}
h^2 = \frac{\sigma_g^2}{\sigma_g^2 + \sigma_e^2} \in [0, 1).
\end{equation*}
In spatially resolved transcriptomics \citep{marx2021method}, when elements of
$Y$ are a gene's expression at $n$ distinct two- or three-dimensional spatial
positions in a tissue sample and $K$ is a matrix encoding distances between
positions, $h^2$ is known as the fraction of spatial variance
\citep{svensson2018spatialde,
kats2021spatialde2,weber2022nnsvg,zeng2022statistical}.
 The parameter $h^2$ is related to $\tau = \sigma^2_g / \sigma^2_e$, which has been the focus of previous research on confidence intervals under \eqref{eq:lmm} \citep[e.g.,][]{Crainiceanu.Ruppert2004}, as $h^2 = \tau / (1 + \tau)$. Because the naming of parameters is
inconsistent in the literature, we refer to all of $\sigma_g^2$, $\tau$, and
$h^2$ as variance components for simplicity. To be concrete, we will focus on
$h^2$ as that is common in the motivating applications, but
similar results often apply in different parameterizations. A further discussion of how our results could be extended to other parameterizations and models, including ones with more variance components or a 
parameterized $K$, is in the Supplementary Material.

Though \eqref{eq:one_varcomp} may seem a simple model, inference on $h^2$ can be
complicated, especially when it is nearly, or equal to, zero or one, as those
are boundary points of the parameter set. It is well-known that common
test-statistics, such as Wald and likelihood ratio statistics, have non-standard
distributions on boundary points. The distributions can often be obtained by
simulation \citep[e.g.,][]{Crainiceanu.Ruppert2004, Schweigeretal2016,
Guedon.etal2024} or approximated, either using asymptotic theory
\citep[e.g.,][]{Self.Liang1987,Geyer1994,
Stram.Lee1994,Stram.Lee1995,Baey.etal2019}, or by other means
\citep[e.g.][]{Wood2013}. Thus, it is often possible to construct tests with
(asymptotically) correct pointwise size, and the likelihood ratio test has
recently been suggested for use in related settings
\citep{Battey.McCullagh2023}. However, creating a reliable confidence interval by
inverting such a test-statistic is more complicated, for at least two reasons. 

First, one in general has to compute the statistic for every relevant $h^2$.
Thus, methods designed specifically for testing $h^2 = 0$ are
inapplicable, and simulation-based methods can be prohibitively time-consuming.
Secondly, even when inverting a test with asymptotically correct pointwise size
is computationally feasible, the resulting confidence region can have poor
coverage properties. Specifically, the actual coverage probability often depends
substantially on how close the true parameter is to the boundary. More formally,
test-statistics with different asymptotic distributions on boundary and interior
points lead to asymptotically {\it incorrect} uniform coverage probability
\citep[Lemma 2.5]{Ekvall.Bottai2022}.

\begin{figure}

\centering
\includegraphics[width=\linewidth]{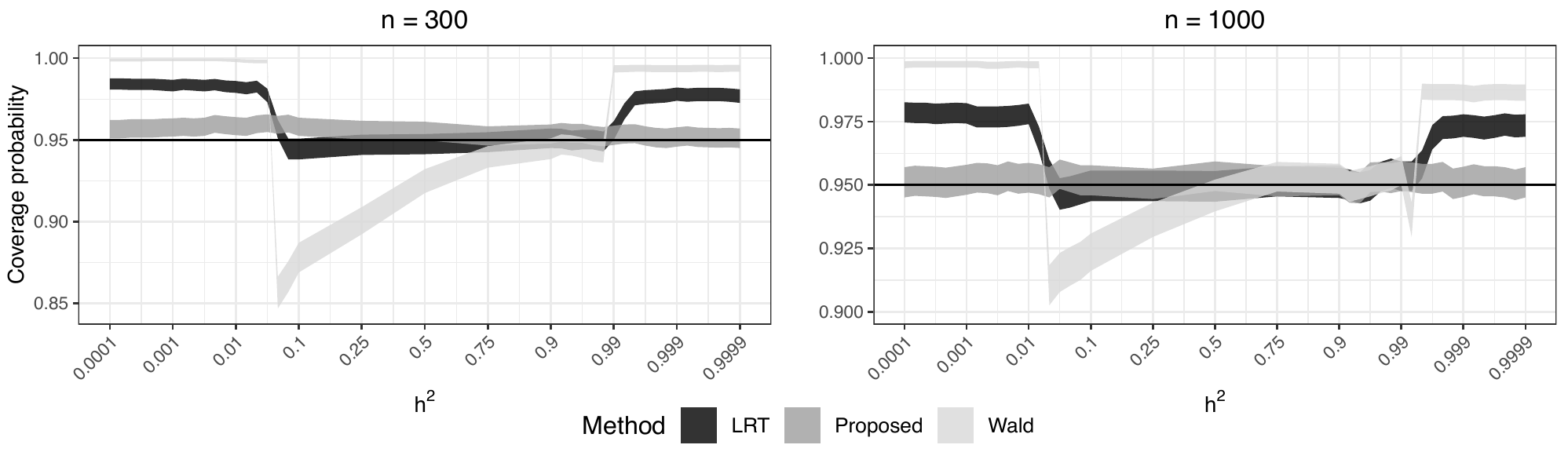}

\caption{Monte Carlo estimates of coverage probabilities for $h^2$ in
    \eqref{eq:one_varcomp} with $\sigma^2_g + \sigma_e^2 = 1$, $\beta = 0$, $X
    \in \R{n\times 5}$ a matrix of independent standard normal entries, and
    $K_{ij} = 0.95^{\vert i - j\vert}$. The solid horizontal line indicates the
    nominal level. The line widths provide 95\% confidence bands for the
    coverage probability based on $10^4$
    trials.}
\label{fig:illustration}
\end{figure}

To illustrate the issues, Fig.~\ref{fig:illustration} shows coverage
probabilities for different values of $h^2$ for Wald, likelihood ratio, and
proposed confidence intervals defined in Section ~\ref{sec:model}. For interior
points of the parameter set, the confidence intervals are obtained by inverting
test-statistics using quantiles of the chi-square distribution with one degree
of freedom, as suggested by classical theory. Notably, the coverage
probabilities for Wald and likelihood ratio intervals are substantially
different from the nominal level when $h^2$ is near zero or one. The reason is
that, when the true parameter is close to the boundary, the distributions of the
test-statistics are close to what they are on the boundary. Consequently, even though the
parameter is interior, classical asymptotic theory for interior points
provides a poor approximation. Removing the boundary points from the parameter
set or using a larger sample size is not sufficient to address the issue, as
Fig.~\ref{fig:illustration} illustrates; loosely speaking, for every $n$, there
are interior points close enough to the boundary to cause problems. Asymptotic
theory for related settings suggests the problematic points are within a
distance of order $n^{-1/2}$ of the boundary
\citep{Rotnitzky.etal2000,Bottai2003}. In the Supplementary Material, we include
a version of Fig.~\ref{fig:illustration} with horizontal axis range $h^2 \in
[.005, .01]$, where it is more clear that the problematic region indeed changes
with $n$.

For reasons to become apparent, we focus on test-statistics based on the score
and information of the restricted likelihood. By doing so we are able to
construct a $(1- \alpha)$ confidence region $\mathcal{R}_n(\alpha)$ that
satisfies, for any compact subset of the parameter set, say $C$, and as $n\to \infty$,
\begin{equation} \label{eq:asy_coverage}
    \sup_{\theta \in C} \left \vert \pr_\theta\{h^2 \in \mathcal{R}_n(\alpha)\} - 
        (1 - \alpha)\right\vert \to 0.
\end{equation}
That is, the proposed confidence region, which is typically an interval, has
asymptotically correct uniform coverage probability on compact sets. This
property leads to near-nominal coverage of every practically relevant $h^2$,
including $h^2$ near or equal to zero, as is illustrated in
Fig.~\ref{fig:illustration}. We emphasize that when focusing on uniform
inference in the presence of boundary points, conclusions about score
test-statistics in general do not carry over to Wald and likelihood ratio
test-statistics. Indeed, our results show, in a sense to be made precise, that
certain score statistics are asymptotically pivotal, but Wald and likelihood
ratio test-statistics are not. The score statistics can therefore be inverted
using the same quantiles for every point in the parameter set, leading to a
confidence interval that is easy to implement and orders of magnitude faster to
compute than competing ones. In a typical example we consider, our confidence
intervals take less than a second to compute, whereas the competitors often take
500 seconds, and sometimes much longer.

Previously, the score standardized by expected information has been shown to
give confidence regions with asymptotically correct uniform coverage probability
in settings with singular Fisher information
\citep{Bottai2003,Ekvall.Bottai2022}. We build on similar ideas here, but the
setting, and hence the arguments required for the main results, are quite
different. In particular, here, the information matrix is positive definite
under weak conditions (Theorem \ref{thm:posdef}). There are also two nuisance
parameters, $\beta$ and $\sigma^2$. While dealing with $\beta$ is relatively
straightforward since $\beta$ and $(h^2, \sigma^2)$ are orthogonal in the Fisher
information sense, our results are, to the best of our knowledge, the first of
their kind with non-orthogonal nuisance parameters. For instance,
\citet{Ekvall.Bottai2022} assume $\sigma^2_e > 0$ is known. Moreover, here, the
number of predictors can grow with the sample size. This is an important
development as the restricted likelihood is typically used precisely when the
number of predictors is smaller than the number of observations, but large
enough to induce substantial bias in the inference on $h^2$.

{\color{blue} }

\section{Model} \label{sec:model}

\subsection{Likelihood functions}

Let $\Sigma = \cov_{\theta}(Y) = \sigma_g^2K + \sigma_e^2I_n$. A log-likelihood
corresponding to \eqref{eq:one_varcomp} is
\begin{equation}\label{eq:likelihood_sigmas}
    l(\beta, \sigma_g^2, \sigma_e^2) = l(\beta, \sigma_g^2, \sigma_e^2; Y, X, K) =
-\frac{1}{2}\left\{\log\vert\Sigma\vert+(Y-X\beta)^\T\Sigma^{-1}(Y-X\beta)\right\},
\end{equation}
the domain of which is the parameter set $\R{p} \times [0, \infty) \times (0,
\infty)$. It is important that \eqref{eq:likelihood_sigmas} allows $\sigma^2_g =
0$, which is often of practical interest. When $K$ is positive definite, points
with $\sigma^2_e = 0$ and $\sigma_g^2 > 0$ can be added to the parameter set, but
with semi-definite $K$, \eqref{eq:likelihood_sigmas} requires
$\sigma_e^2 > 0$. Assuming $X$ has full column rank, the restricted
log-likelihood, often used for inference on the covariance parameters, is
\[
    l_R(\sigma_g^2, \sigma_e^2) = -\frac{1}{2}\{\log \vert \Sigma \vert + 
    \log \vert X^\T \Sigma^{-1}X\vert + (Y - X\tilde{\beta})^\T \Sigma^{-1}
    (Y - X\tilde{\beta})\},
\]
where $\tilde{\beta} = \tilde{\beta}(\sigma_g^2, \sigma_e^2) = (X^\T
\Sigma^{-1}X)^{-1}X^\T \Sigma^{-1} Y$ is the partial maximizer of $l(\beta,
\sigma^2_g, \sigma^2_e)$ in $\beta$, for fixed $\sigma_g^2$ and $\sigma^2_e$.
The domain of $l_R$ is $[0, \infty) \times (0, \infty)$. The assumption that $X$
have full column rank is for notational convenience; one could define $l_R$
similarly as long as the column rank is less than $n$. Specifically, if the
column rank of $X$ is $p'< n$ and $V \in \R{n\times (n - p')}$ is a
semi-orthogonal matrix satisfying $V^\T X = 0$, then the restricted likelihood is
the likelihood of $V^\T Y$ \citep{Harville1974}.

By spectral decomposition, $K = O\Lambda O^\T$ for some orthogonal $O \in
\R{n\times n}$ and $\Lambda = \diag(\lambda_1, \dots, \lambda_n)$; here and
elsewhere, eigenvalues are in decreasing order unless otherwise stated. It
follows that $\Sigma = O (\sigma^2_g\Lambda + \sigma^2_e I_n)O^\T$ and, hence,
\[
    l(\beta, \sigma^2_g, \sigma^2_e) = -\frac{1}{2}
        \sum_{i = 1}^n \left\{\log(\lambda_i\sigma^2_g + \sigma^2_e) + 
        \frac{(O^\T Y - O^\T X \beta)_i^2}{\lambda_i \sigma^2_g  + \sigma^2_e}
        \right\},
\]
where $(\cdot)_i$ means the $i$th element; and 
\[
    l_R(\sigma^2_g, \sigma^2_e) = l(\tilde{\beta}, \sigma^2_g, \sigma^2_e) - \log
    \vert (O^\T X)^\T (\sigma^2_g \Lambda + \sigma^2_e I_n)^{-1} O^\T X\vert / 2.
\] 
Thus, when developing theory, we may upon replacing $Y$ and $X$ by $O^\T Y$ and
$O^\T X$, respectively, assume diagonal covariance matrices $K = \Lambda$
(equivalently, $O = I_n$) and $\Sigma = \sigma^2_g \Lambda + \sigma^2_e I_n$
without loss of generality. When doing so, the observations are ordered so that
the $i$th response $Y_i$ corresponds to the $i$th eigenvalue $\lambda_{i}$.

At the expense of more notation, the assumption on $\Sigma$ could be relaxed to
$\Sigma = \sigma^2_g K + \sigma^2_e W$ for some known positive definite $W \in
\R{n\times n}$. Indeed, in that case $Y$, $X$, and $K$ can be replaced by,
respectively, $W^{-1/2}Y$, $W^{-1/2}X$, and $W^{-1/2}K W^{-1/2}$ before
repeating the above argument to get a diagonal $\Sigma$.

In several motivating applications, the parameter of interest is $h^2 =
\sigma^2_g / (\sigma^2_g + \sigma^2_e)$. Therefore, we consider a common
reparameterization in terms of $h^2$ and $\sigma^2 = \sigma^2_g + \sigma^2_e$.
Then, $\Sigma = \sigma^2 O\{h^2\Lambda + (1 - h^2)I_n\}O^\T$. Overloading the
letter $l$ and assuming $O = I_n$ without loss of generality, the log-likelihood
becomes
\begin{equation} \label{eq:loglik}
    l(\beta, h^2, \sigma^2) = -\frac{1}{2}
    \sum_{i = 1}^n \left\{\log(\sigma^2) + \log(\lambda_i h^2 + 1 - h^2) + 
    \frac{(Y_i - X_i^\tsp \beta)^2}{\sigma^2(\lambda_i h^2 + 1 - h^2)}
    \right\},
\end{equation}
where $X_i^\tsp$ is the $i$th row of $X$. The domain of $l$ is the parameter set
$\R{p} \times [0, 1) \times (0, \infty)$. The corresponding restricted
log-likelihood, with domain $[0, 1) \times (0, \infty)$, is
\[
    l_R(h^2, \sigma^2) = l(\tilde{\beta}, h^2, \sigma^2) +
    \frac{p}{2}\log (\sigma^2) - \frac{1}{2}\log \vert X^\T 
    \{h^2\Lambda + (1 - h^2)I_n\}^{-1}X\vert.
\]

\subsection{Test-statistics and confidence regions}\label{sec:test-statistics}

There are many potential test-statistics for inference based on $l(\beta, h^2,
\sigma^2)$ and $l_R(h^2, \sigma^2)$. However, most common ones, like Wald and
likelihood ratio statistics, do not give confidence regions satisfying
\eqref{eq:asy_coverage} in boundary settings in general. Therefore, we focus on
score statistics standardized to have mean zero and diagonal covariance matrix.
Such test-statistics have previously been shown to lead to
results like \eqref{eq:asy_coverage} in settings where, unlike here, the Fisher information
is singular \citep{Ekvall.Bottai2022}. Because our goal is to
construct confidence regions for the covariance parameters, effectively treating
$\beta$ as a nuisance parameter, we primarily work with the restricted
likelihood.

Let $V_1 = V_1(h^2, \sigma^2) = \partial \Sigma /
\partial h^2 = \sigma^2O(\Lambda-I_n)O^\T$ and $V_2 = V_2(h^2, \sigma^2) =
\partial \Sigma / \partial \sigma^2 = O(h^2\Lambda + (1 - h^2)I_n)O^\T$. Similarly,
let $U_1(h^2, \sigma^2) = \partial l_R(h^2, \sigma^2) / \partial h^2$ and
$U_2(h^2, \sigma^2) = \partial l_R(h^2, \sigma^2) / \partial \sigma^2$, so
the restricted score is $U(h^2, \sigma^2) = [U_1(h^2, \sigma^2), U_2(h^2,
\sigma^2)]^\T$. Routine calculations show, for $j\in \{1, 2\}$,
\begin{align*}
    U_j(h^2, \sigma^2) &=
    \frac{1}{2}(Y - X\tilde{\beta})^\T \Sigma^{-1}V_j \Sigma^{-1} (Y - X\tilde{\beta}) 
    -\frac{1}{2}\tr(\Sigma^{-1}V_j)\\
    &\quad + \frac{1}{2}\tr(\Sigma^{-1}X(X^\T\Sigma^{-1}X)^{-1}X^\T\Sigma^{-1}V_j).
\end{align*}
Observe $U_j$ is defined on $[0, 1)\times(0, \infty)$ and has a continuous
extension to points where $h^2 = 1, \sigma^2 > 0$ if $K$ is positive definite. The expression
simplifies substantially for $j = 2$ since $V_2 = \Sigma / \sigma^2$.
Specifically, assuming $O = I_n$,
\begin{equation}\label{eq:score_sigma2p}
    U_2(h^2, \sigma^2) = -\frac{n - p}{2 \sigma^2} +
    \frac{1}{2 \sigma^4}\sum_{i = 1}^n \frac{(Y_i - X_i^\T\tilde{\beta})^2}
    {h^2 \lambda_i + 1 - h^2}.
\end{equation}
Let $\mathcal{I}(h^2, \sigma^2)$ denote the restricted information matrix, that
is, the covariance matrix of $U(h^2, \sigma^2)$ when $h^2$ and $\sigma^2$ are
the true parameters. We next give an expression for $\mathcal{I}(h^2, \sigma^2)$
for easy reference. The proof, along with those of other formally stated
results, is in the Supplementary Material. 

Let $P = \Sigma^{-1/2}X(X^\T\Sigma^{-1}X)^{-1}X^\T\Sigma^{-1/2}$, $Q = I_n - P$,
$D = \{h^2 \Lambda + (1 - h^2)I_n \}^{-1}(\Lambda - I_n)$, and $H = ODO^\tsp$.

\begin{proposition} \label{prop:finf}
    If \eqref{eq:one_varcomp} holds for some $X$ with full column rank, then the
    restricted information matrix has elements
    \begin{align*}
        \mathcal{I}_{11}(h^2, \sigma^2) = \frac{1}{2}\tr(Q H QH); ~~
        \mathcal{I}_{12}(h^2, \sigma^2) = \frac{1}{2\sigma^2}\tr(Q H);~~
        \mathcal{I}_{22}(h^2, \sigma^2) & = \frac{n - p}{2\sigma^4}.
    \end{align*}
\end{proposition}

For inference on $(h^2, \sigma^2)$ jointly, we consider the (quadratic)
restricted score standardized by the restricted information:
\begin{equation} \label{eq:test_stat}
    T^R_n(h^2, \sigma^2) = U(h^2, \sigma^2)^\T
    \mathcal{I}^{-1}(h^2, \sigma^2)U(h^2, \sigma^2),
\end{equation}
For inference on $h^2$ only, we use
\begin{equation} \label{eq:test_stat_h}
    T^R_n(h^2) = U_1(h^2, \tilde{\sigma}^2)^2
    \mathcal{I}^{11}(h^2, \tilde{\sigma}^2),
\end{equation}
where $ \mathcal{I}^{11}(h^2, \sigma^2)$ is the leading element of $
\mathcal{I}^{-1}(h^2, \sigma^2)$ and $\tilde{\sigma}^2 = \tilde{\sigma}^2
(h^2)$ is the partial maximizer of $l_R(h^2, \sigma^2)$ in $\sigma^2$, for a
fixed $h^2$. By \eqref{eq:score_sigma2p}, assuming $O = I_n$,
\begin{equation} \label{eq:sigma_p2_hat}
    \tilde{\sigma}^2 = \frac{1}{n - p}\sum_{i = 1}^n 
    \frac{(Y_i - X_i^\T\tilde{\beta})^2} {h^2 \lambda_i + 1 - h^2}  = \frac{1}{n - p}(Y - X\tilde{\beta})^\T \{h^2 \Lambda + (1 - h^2)I_n\}^{-1} (Y - X\tilde{\beta}).
\end{equation}

When $X$ has column rank less than $n$, $\tilde{\sigma}^2$ is greater than
zero with probability one, and hence a valid partial maximizer. Note
$\tilde{\beta}$ does not depend on $\sigma^2$, so the last display indeed
gives an explicit expression for $\tilde{\sigma}^2$. For inference on $h^2$,
we sometimes also consider $S_n^R(h) = U_1(h^2, \tilde{\sigma}^2)
\mathcal{I}^{11}(h^2, \tilde{\sigma}^2)^{1/2}$, that is, the signed square root
of \eqref{eq:test_stat_h}.

The proposed test-statistics require a positive definite $\mathcal{I}(h^2,
\sigma^2)$, which the following result establishes. Notably,
$\mathcal{I}^{11}(h^2, \sigma^2)$ is positive if $\mathcal{I}(h^2, \sigma^2)$ is
positive definite. To understand the next result, notice that
$Q$ has rank $n-p$, so $QHQ$ has at most $n-p$ non-zero eigenvalues, and recall
$V \in \R{n \times (n - p)}$ is a semi-orthogonal matrix such that $V^\T X =
0$.

\begin{theorem} \label{thm:posdef}
   The information matrix $\mathcal{I}(h^2, \sigma^2)$ is singular if and only
   if the following equivalent conditions hold: (i) the $n-p$ possibly non-zero
   eigenvalues of $QHQ$ are identical; (ii) $V^\T K V = c I_{n-p}$ for some $c\geq 0$.
\end{theorem}

The proof of Theorem \ref{thm:posdef}
is relatively straightforward using the expressions in Proposition \ref{prop:finf}. 
The key observation is that the determinant of $\mathcal{I}(h^2, \sigma^2)$
is
\[
    \frac{1}{2\sigma^4}\left\{ (n-p)\tr(QHQ QHQ) - \tr(QHQ)^2\right\},
\]
which is strictly positive by Jensen's inequality applied to the $n-p$ possibly
non-zero eigenvalues of $QHQ$ unless they are all identical, in which case the
determinant is zero.

Theorem \ref{thm:posdef} is particularly intuitive if $p = 0$, so that $Q = V=
I_n$. Then the eigenvalues of $QHQ$ are the diagonal elements of $D$, which are
identical if and only if the diagonal elements of $\Lambda$ are; that is, if and
only if $\Lambda$, and hence $K$, is proportional to the identity. The
parameters are unidentifiable for such $K$, so we will generally assume $K$ is
not proportional to the identity, leading to a positive definite information
matrix.

By inverting the proposed test-statistics using quantiles from chi-squared distributions, we get the confidence regions
\begin{align}\label{eq:confreg}
\begin{split}
    \mathcal{R}_n(\alpha) &= \{h^2 \in [0, 1] : T_n^R(h^2) \leq q_{1 - \alpha, 1}\};\\
    \mathcal{R}^2_n(\alpha) &= \{(h^2, \sigma^2) \in [0, 1] \times (0, \infty) : T_n^R(h^2, \sigma^2) \leq q_{1 - \alpha, 2}\}.
\end{split}
\end{align}
 When $h^2 = 1$ is not in the parameter set, we take $T^R_n(1) = T^R_n(1, \sigma^2) = \infty$ so the confidence regions are well defined.
 
\section{Asymptotic distributions} \label{sec:asy_theory}

We assume $K_n = \Lambda_n = \diag(\lambda_{n1}, \dots, \lambda_{nn})$ in this
section without loss of generality, and hence $\Sigma_n = \sigma^2_{n}\{(1 - h_n)I_n +
h_n \Lambda_n\}$ and $H_n = D_n$ are diagonal. Here, a subscript $n$ has been added
to $\sigma^2$ and other quantities since we will consider a sequence
$\{(\beta_n, h^2_n, \sigma^2_{n})\}$ of parameters indexed by the sample size,
and a corresponding sequence of models. Similarly, the number of predictors $p =
p_n$ can change with $n$. By considering a sequence of (true) parameters, we can
obtain the asymptotic distribution of $T^R_n(h^2_n, \sigma^2_n)$ and
$T_n^R(h_n^2)$ under that sequence. This is useful because
\eqref{eq:asy_coverage} holds for $\mathcal{R}_n(\alpha)$ in \eqref{eq:confreg} if and only if the asymptotic distribution of
$T_n^R(h_n^2)$ is the same under any convergent sequence of parameters
\citep[Lemma 2.5]{Ekvall.Bottai2022}, and similarly for $\mathcal{R}^2_n(\alpha)$. This result essentially follows from the
equivalence of continuous convergence and uniform convergence on compact sets,
applied to the sequence of functions defined by $\theta \mapsto
\pr_{\theta}\{T_n^R(h^2) \leq q_{1-\alpha, 1}\}$. More generally, examining the distribution under a sequence of
parameters tending to a point, can lead to a better understanding of the
behavior of the test-statistic near that point. Thus, for our purposes,
sequences tending to boundary points are of particular interest. The next result
is our first along these lines.

To state the result, let $\gamma_i(\cdot)$ denote the $i$th eigenvalue in
decreasing order.

\begin{lemma} \label{lem:asy_dist}
    Assume that, for every $n \in \{1, 2, \dots\}$, $Y$ satisfies
    \eqref{eq:one_varcomp} with an $X$ with full column rank, $K_n = \Lambda_n$,
    and parameters $(\beta_n, h^2_n, \sigma^2_{n})$. If, as $n\to \infty$, (i) $p/n \to 0$, (ii) $\sum_{i = 1}^p \gamma_i(D_n^2) / \tr(D_n^2) \to 0$, (iii) $\limsup_{n\to \infty} \{n^{-1}\tr(D_n)^2 / \tr(D_n^2)\} < 1$; then, in distribution,
    \begin{equation} \label{eq:conv_dist}
    T_n^R(h_n^2, \sigma^2_{n})\to \chi_2^2.
    \end{equation}
\end{lemma}

The proof of Lemma \ref{lem:asy_dist} establishes \eqref{eq:conv_dist} by
showing asymptotic bivariate normality of $\mathcal{I}_n(h_n^2,
\sigma_n^2)^{-1/2}U_n(h_n^2, \sigma^2_n)$. The key observation is that the
elements of $U_n(h_n^2, \sigma^2_n)$ are quadratic forms in multivariate normal
vectors, centered to have mean zero. Therefore, their distributions are those of
linear combinations of independent chi-square random variables. The conditions
of the theorem ensure Lyapunov's central limit theorem applies to those
linear combinations.

Conditions (ii) and (iii) concern $\Lambda_n$ and $h_n^2$, but neither of
$\beta_n$ and $\sigma^2_{n}$ appears in the conditions. Thus,
\eqref{eq:conv_dist} can hold for sequences one may not expect, such as
those where $\sigma^2_{n} \to 0$. This suggests the behavior of our
test-statistic is insensitive to the true $\sigma^2$, and may perform well
even if $\sigma^2$ is near the boundary, that is, near zero.

Condition (ii) says the sum of the $p$ greatest eigenvalues of $D_n^2$ should
be negligible compared to the sum of all its eigenvalues. An eigenvalue, or
diagonal element, of $D_n^2$ will be relatively large if either (a) the
corresponding element of $\Lambda_n$ is relatively close to zero and $h^2$
is close to one, or (b) the corresponding element of $\Lambda_n$ is
relatively large. Thus, condition (ii) suggests the test-statistic may not
behave well if a few elements of $O^\T Y$ have much
larger variance than all other ones, or if some elements have very small
variance and most of it is attributable to $K$.

Condition (iii) essentially says that, when comparing the squared average of
the eigenvalues of $D_n$ to the average of the squared eigenvalues, Jensen's
inequality is asymptotically strict; there has to be non-negligible
variation in the diagonal elements of $D_n$. This is intuitive as, if all
the diagonal elements of $D_n$ are the same, then $\Lambda_n$ is
proportional to the identity and $h_n^2$ unidentifiable, as noted in the
remark following Theorem \ref{thm:posdef}. 

Conditions (ii) and (iii) could be replaced by slightly weaker ones by adding
assumptions on how $X$ relates to $K$, but that would substantially increase
notational burden.

Our next result shows that the proposed test-statistic for $h^2$ has the desired
asymptotic distribution under the conditions of Lemma \ref{lem:asy_dist}.
Substantial work is needed to establish the result since $\sigma^2$ is a
nuisance parameter for the purposes of inference on $h^2$ only. In particular,
the test-statistic for $h^2$ cannot depend on or be evaluated at the true
parameter, or null hypothesis, which was possible in Lemma \ref{lem:asy_dist}.

\begin{lemma} \label{lem:asy_dist_h2}
Under the conditions of Lemma \ref{lem:asy_dist}, $T_n^R(h_n^2) \to \chi^2_1$ in distribution as $n\to\infty$.
\end{lemma}

Lemma \ref{lem:asy_dist_h2} is proven by showing that $T_n(h_n^2)$ has the same
asymptotic distribution as $\{U_{n1} - \mathcal{I}_{n12} U_{n2} /
\mathcal{I}_{n22}\}^2 \mathcal{I}_n^{(11)}$, with all quantities evaluated at
the true $(h_n^2, \sigma^2_n)$ and $\mathcal{I}_n^{(11)}$ denoting the leading
element of $\mathcal{I}_n^{-1}$. This result is consistent with what is often
obtained for nuisance parameters in classical settings, with fixed interior
parameters \citep[Section 9.3]{Cox.Hinkley2000}. This is remarkable because those results in
general do not apply under sequences of parameters or with boundary points.

Our next result gives intuitive conditions on model parameters and $K$ which
ensure the conditions in Lemmas \ref{lem:asy_dist} and \ref{lem:asy_dist_h2}
hold. The conditions are not necessary in general, but they are more transparent, and
easier to assess in practice, than those in the lemmas.

\begin{theorem}\label{thm:asy_conv_bivariate}
    Assume that, for every $n \in \{1, 2, \dots\}$, $Y$ satisfies
    \eqref{eq:one_varcomp} with an $X$ with full column rank and parameters
    $(\beta_n, h^2_n, \sigma^2_{n})$. If (i) $\lim_{n\to \infty} p_n/n = 0$; (ii$^*$) $\limsup_{n\to\infty}\lambda_{n1} < \infty $; (iii$^*$) $\liminf_{n\to\infty}\lambda_{nn} > 0$ or
        $\limsup_{n\to\infty}h^2_n < 1$; (iv$^*$) there exists an $\epsilon > 0$ such that for any $c \in
        [0,\infty)$, with $k^c_n$ denoting the number of eigenvalues of
        $\Lambda_n$ in $[c-\epsilon, c + \epsilon]$, it holds that
        $\limsup_{n\to \infty }k_n^c / n < 1$,
    then, in distribution as $n\to \infty$,
    \begin{equation}
        T_n^R(h_n^2, \sigma^2_{n}) \to \chi_2^2; ~~T_n^R(h_n^2)\to \chi_1^2.
    \end{equation}
\end{theorem}

Since $\Lambda_n$ is known, condition (ii$^*$) can be made to hold, for example by
replacing $K$ by $K/\lambda_{n1}$. Then, for the conclusions to hold, conditions (iii$^*$) and (iv$^*$) would need to hold for the new eigenvalues $\lambda_{ni}/\lambda_{n1}$,
$i \in \{1, \dots, n\}$.  Condition (iii$^*$) illustrates that if $K$ is full rank, then
reliable inference on $h^2$ near one is possible; this is formalized further in
Corollary \ref{corol:coverage}. Condition (iv$^*$) prevents the eigenvalues of $K$ from
concentrating in some neighborhood asymptotically. That is, it prevents $K$ from
being essentially proportional to the identity asymptotically.

\begin{corollary} \label{corol:coverage}
    Under condition (i), (ii$^*$), and (iv$^*$) of Theorem \ref{thm:asy_conv_bivariate}, $\mathcal{R}_n(\alpha)$ and $\mathcal{R}^2_n(\alpha)$ in \eqref{eq:confreg} have asymptotically correct uniform coverage probability on
    compact subsets of $[0, 1)$ and $[0, 1)\times (0, \infty)$, respectively
    (c.f. \eqref{eq:asy_coverage}). The conclusion can be strengthened to
    compact subsets of $[0, 1]\times (0, \infty)$ and $[0, 1]$, respectively, if
    the first part of condition (iii$^*$) of Theorem \ref{thm:asy_conv_bivariate} also
    holds.
\end{corollary}

\section{Computational considerations}

In this section, we consider the computational aspects of our test-statistics.
Like popular competing methods, we require a one-time preprocessing step where
the eigendecomposition of $K = O\Lambda O^\T$ is used to construct $Y_O = O^\T Y$
and $X_O = O^\T X$. In some applications, the same $K$ is used for many
different choices of response vector $Y$, with separate inference on $h^2$ for
each response vector. Then, the pre-processing step is only performed once.

We show that, after the pre-processing, the number of floating point operations
required to evaluate the proposed test-statistics grows linearly in $n$, making
our methods feasible for large-scale applications, such as those described in
Section \ref{sec:Introduction}. More specifically, we show that, if $p$ is
potentially growing with $n$, the number of operations required is $O(np^2 +
p^3)$.

Recall $V_2 = h^2\Lambda + (1 - h^2)I_n$ and define $\tilde{P} =
V_2^{-1/2}X_O(X_O^\T V_2^{-1}X_O)^{-1}X_O^\T V_2^{-1/2}$, $\tilde{Q} = I_n -
\tilde{P}$, $\tilde{\beta} = (X_O^\tsp V_2^{-1}X_O)^{-1}X_O^\tsp V_2^{-1}Y_O$, and
$\tilde{R} =Y_O - X_O\tilde{\beta}$. Let $\tilde{R}_{i}$ denote the $i$th
element of $\tilde{R}$ and $ Q_{ii}$ the $i$th diagonal element of $Q$.
The score function is then
\begin{equation}\label{eq:simple_score}
    \mathit{U}(h^2, \sigma^2) = \frac{1}{2}\begin{pmatrix}
        \sum_{i=1}^n D_{ii}\left(\frac{\sigma^{-2}\tilde{R}_i^2 }{h^2 \lambda_{i} + 1 - h^2 } -  \tilde{Q}_{ii}\right) \\
        -\frac{n - p}{ \sigma^2} +
    \frac{1}{\sigma^4}\sum_{i= 1}^n \frac{\tilde{R}_i^2}
    {h^2 \lambda_{i} + 1 - h^2}
         \end{pmatrix}.
\end{equation}
Both terms in \eqref{eq:simple_score} involve the summation of $n$ terms.
Construction of $\tilde{R}$ requires $O(np)$ operations, and computation of all
$\tilde{Q}_{ii}$ requires $O(np^2)$ operations. We do not
compute or store the entire matrix $\tilde{Q}$. 

To compute the information efficiently, note $Q = O\tilde{Q}O^\tsp$, so by
Proposition \ref{prop:finf}, 
$$ \mathcal{I}(h^2, \sigma^2) = \frac{1}{2\sigma^4}\left(\begin{array}{cc}
\sigma^4 {\rm tr}(\tilde{Q}D \tilde{Q}D) & \sigma^2 {\rm tr}(\tilde{Q}D)\\
\sigma^2 {\rm tr}(\tilde{Q}D) & n - p\end{array}\right).$$ Because $ {\rm
tr}(\tilde{Q}D) = \sum_{i=1}^n \tilde{Q}_{ii} D_{ii}$, $\mathcal{I}_{12}$ can be
evaluated in $O(n)$ operations using the $\tilde{Q}_{ii}$ computed before.
Notice $2\hspace{1pt}\mathcal{I}_{11} = {\rm tr}(\tilde{Q}D \tilde{Q}D) =
{\rm tr}(D^2) - 2{\rm tr}(\tilde{P}D^2) + {\rm tr}(\tilde{P}D\tilde{P}D)$.
Because $D$ is a diagonal matrix, evaluating ${\rm tr}(D^2)$ requires $O(n)$
operations. Similarly, ${\rm tr}(\tilde{P}D^2) = \sum_{i=1}^n (1 -
\tilde{Q}_{ii})D_{ii}^2$, so evaluating this term also requires only $O(n)$
operations.  Finally, by the cyclic property of the trace operator, ${\rm
tr}(\tilde{P}D\tilde{P}D) = {\rm tr}(\tilde{A} \tilde{B} \tilde{A} \tilde{B})$
where $\tilde{A} = (X_O^\T V_2^{-1}X_O)^{-1} \in \mathbb{R}^{p \times p}$ and
$\tilde{B} = X_O^\T V_2^{-1/2}D  V_2^{-1/2}X_O \in \mathbb{R}^{p \times p}.$ Thus,
evaluating ${\rm tr}(\tilde{A} \tilde{B} \tilde{A} \tilde{B})$ requires $O(np^2
+ p^3)$ operations, with the $p^3$ coming from a Cholesky (or other)
decomposition of $\tilde{A}^{-1}$ to deal with the inverse, and the $np^2$ is
from matrix multiplications.

We have created an R package \texttt{lmmvar} for computing our test-statistics
and confidence intervals, available for download at
\texttt{github.com/yqzhang5972/lmmvar}. The package also provides an implementation of the method by \citet{Crainiceanu.Ruppert2004}.

\begin{figure}
\centering
\includegraphics[width = \textwidth]{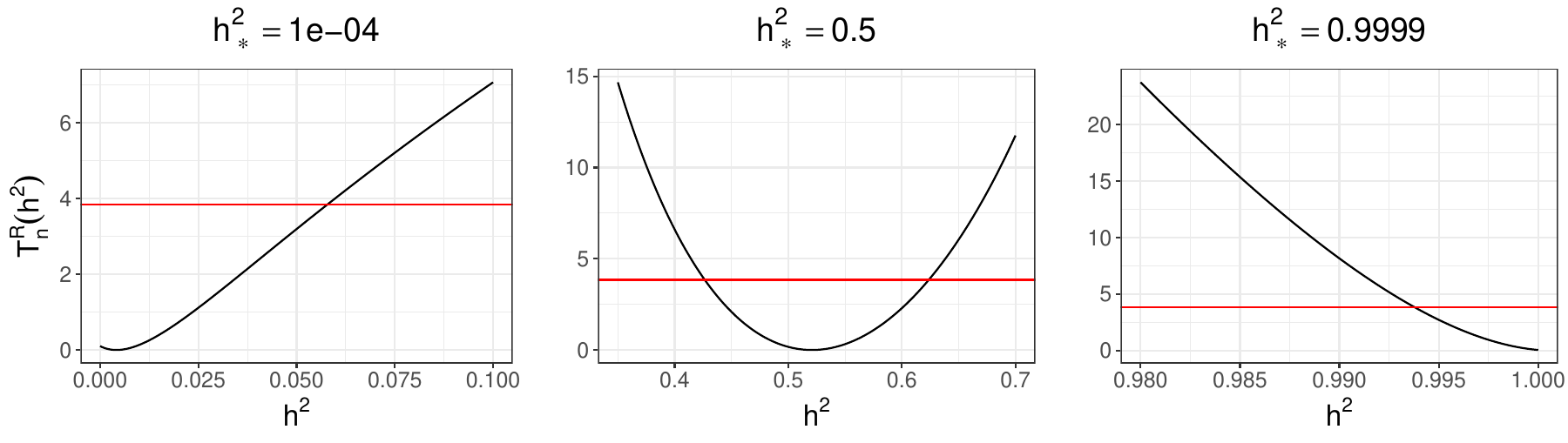}

\caption{The value $T_n^R(h^2)$ with $n = 1000$, data generated under the same
    model as in Fig.~\ref{fig:illustration}, and the true $h^2$, here denoted
    $h_*^2$, taking values $0.0001, 0.5$, and $0.999.$ Horizontal red lines
    indicate the $95$th percentile of the $\chi^2_1$ distribution.}
\label{fig:quasi_convex_whole}
\end{figure}

A confidence interval for $h^2$ can be obtained by evaluating $T_n^R(h^2)$
from at equally spaced points in $[0,1)$ and comparing the values to quantiles
from the chi-square distribution with one degree of freedom. In practice,
however, it is typically possible, and faster, to do a search for the points
where $T_n^R(h^2)$ crosses the desired quantile. More specifically, when
$T_n^R(h^2)$ is strictly quasiconvex in $h^2$, then $\mathcal{R}_n(\alpha)$ is
an interval, so we need only search for points $a$ such that $T_n^R(a) = q_{1,
1-\alpha}$. While we have not been able to prove $T_n^R(h^2)$ is quasiconvex in
general due to technical difficulties, we have not encountered an example where
it is not. Typically, $T_n^R(h^2)$ looks roughly like one of the three cases in
Fig.~\ref{fig:quasi_convex_whole}. 
Thus, in our software we first find a point $a'$ such that $T_n^R(a') < q_{1,
1-\alpha}$.  More specifically, we start a ternary search for the minimum of
$T_n^R$ and terminate the first time a value less than $q_{1, 1-\alpha}$ is
encountered. Then, under quasiconvexity and assuming $a'$ is not on the boundary, the intervals $[0,T_n^R(a')]$ and
$[T_n^R(a'),1)$ contain lower and upper bounds of the interval
$\mathcal{R}_n(\alpha)$, respectively. We then use a bisection search to find $b
\in [0,T_n^R(a')]$ such that $T_n^R(b) = q_{1, 1-\alpha}$, and similarly for
$[T_n^R(a'),1)$. If the point $a'$ is on the boundary, as in the first and third
plot of Fig.~\ref{fig:quasi_convex_whole}, only one bisection search is needed.
In either case, this procedure often requires less computing time than a
grid search.

We have also found that, in practice, $S_n^R(h^2) = \{\mathcal{I}_n^{11}(h^2,
\tilde{\sigma}^2)\}^{1/2}U_1(h^2, \tilde{\sigma}^2)$ is typically monotonely
decreasing in $h^2$, so a search can also be based on that statistic and
comparisons to standard normal quantiles; see Section \ref{sec:gene_ranking} for
an example with one-sided confidence intervals.

\section{Numerical studies}

\subsection{Finite-sample interval width and coverage probabilities}

To supplement our asymptotic theory, we study how $n$, $h^2$, and the
eigenvalues of $K$ affect the coverage probability and width of our confidence
intervals. We consider $K_{ij} = \rho^{|i-j|}$ for $\rho \in (0, 1)$ and $i, j
\in \{1, \dots, n\}$. When $\rho$ is close to one, $K$ is approximately
$1_n1_n^\T $, where $1_n$ is an $n$-vector of ones. Thus, approximately,
$\lambda_{n1} = n$ and $\lambda_{ni} = 0$, $i \geq 2$, which is inconsistent
with conditions (ii$^*$) and (iv$^*$) of Theorem \ref{thm:asy_conv_bivariate}. Conversely,
when $\rho$ is close to zero, $K$ is approximately an identity matrix, which is
inconsistent with condition (iv$^*$) of Theorem \ref{thm:asy_conv_bivariate}. Thus, we
expect our test-statistic to perform best when $\rho$ is not too close to either
zero or one.

\begin{figure}
\centering
\includegraphics[width=\linewidth]{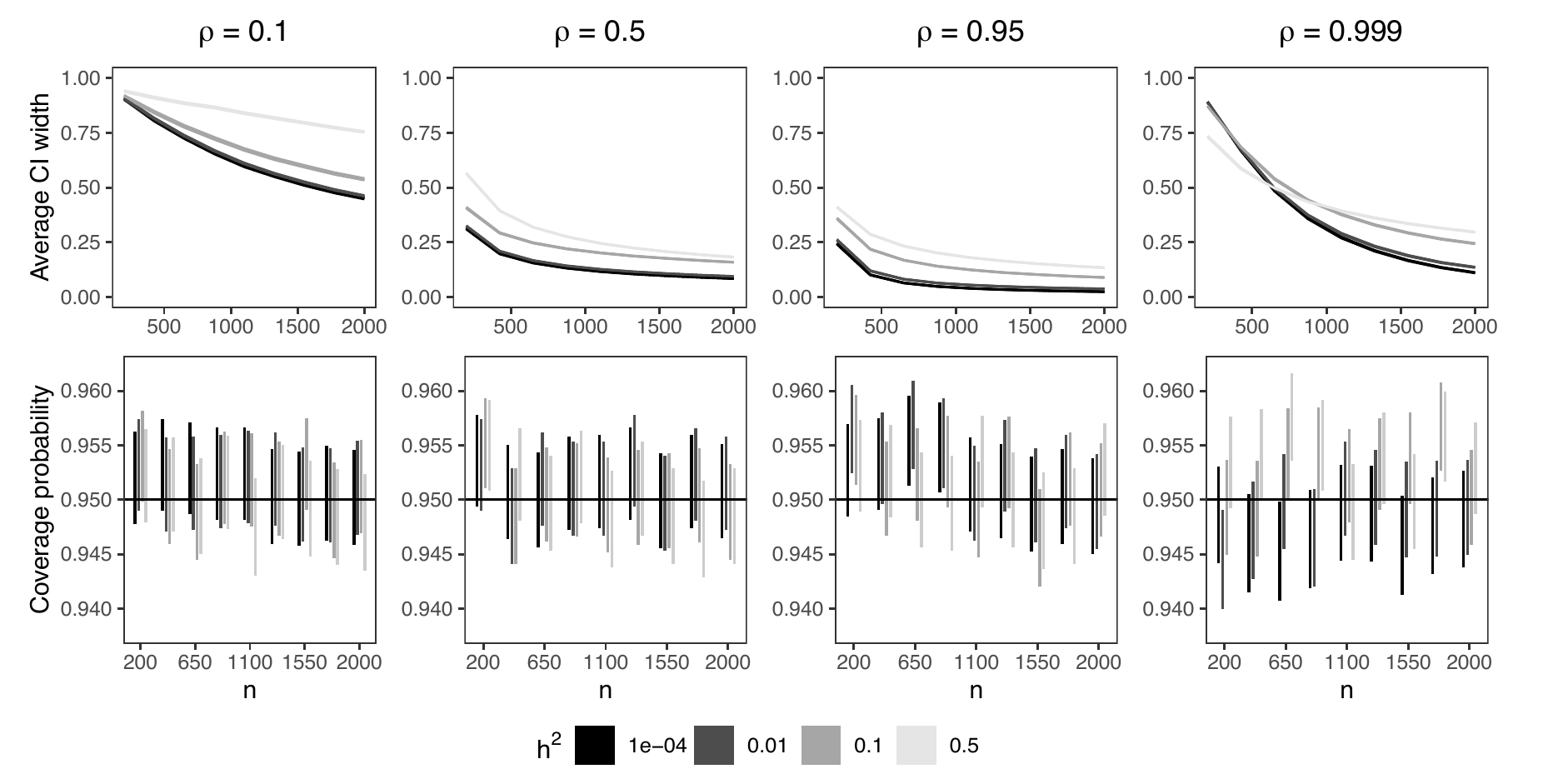}
\caption{(Top panel) The average width of our confidence interval for $h^2$ and (bottom panel) the average coverage probability  as $n$ varies with $\rho \in \{0.1, 0.5, 0.95, 0.999\}$. In the top panels, the widths of the lines provide 95\% confidence bands; and in the bottom panel, the bars contain the 95\% confidence intervals, each based on $10^4$ trials.}
\label{fig:EigenSims}
\end{figure}

In each setting considered, we generate $X \in \mathbb{R}^{n \times 5}$ with
independent and identically distributed standard normal entries, and then
generate $Y \sim \mathcal{N}_n(X\beta, h^2 K + ( 1- h^2) I_n)$ (i.e.,  $\sigma^2
= 1$) with $\beta = 0$. For every combination of the triplet $(n, h^2, \rho)$,
we generate $10^4$ independent realizations of $Y$ and compute the corresponding
95\% confidence intervals for $h^2$ for each realization. Thus, coverage probability and average width intervals in Figure.~\ref{fig:EigenSims} are estimated based on $10^4$ independent replicates. 

The bottom panel of Fig.~\ref{fig:EigenSims} displays coverage probabilities
with $\rho$ and $n$ varying for four different values of $h^2$. For $\rho \in
\{0.1. 0.5. 0.95\}$, our interval appears to have the coverage probability 0.95
or higher. Only when $\rho = 0.999$ is the evidence of under-coverage for some
combinations of $n$ and $h^2$. 

The top panel of Fig.~\ref{fig:EigenSims} displays interval widths for different
values of $\rho$ and $h^2$. When $\rho = 0.1$, $K \approx I_n$, which means
$h^2$ is approximately unidentifiable. Consequently, interval widths are too
large to be practically useful, despite having the nominal coverage level. With
$\rho = 0.5$ or $0.95$, the widths are reasonably small and decrease as $n$
increases. When $\rho = 0.999$, the eigenvalues of $K$ are approximately $(n, 0,
\dots, 0)$, which appear to lead to longer confidence intervals. Nevertheless,
as $n$ increases, interval widths decrease as expected. As $n$ approaches 2000,
the widths are small enough to be useful in practice. 

\subsection{Comparison to simulation-based test-statistics}\label{subsec:comparison}

Existing methods for constructing confidence intervals for $h^2$ often rely on
simulation. For example, \citet{Schweigeretal2016} uses the parametric bootstrap
to approximate the distribution of the restricted maximum likelihood estimator of $h^2.$ In comparison to
a naive implementation of the parametric bootstrap, their implementation is very
efficient. Specifically, to estimate the distribution of the estimator when
the true heritability is $h^2$, their method requires (i) drawing $N$
independent samples from \eqref{eq:one_varcomp} with $\sigma^2 = 1$ and null
hypothesis value $h^2$ given;  and (ii) for each sample, finding an interval
containing a local maximizer of the restricted log-likelihood. Since each sample
is needed for step (ii), \citet{Schweigeretal2016} show that it is sufficient to
draw samples from $\mathcal{N}_n(0, I_n)$ in step (i) and modify step (ii)
accordingly. The modified step (ii) is performed by evaluating the derivative
of the restricted log-likelihood at fixed grid points. However, for a given
$h^2$ the derivative must, in general, be evaluated at many grid points; and
this procedure must be repeated for all $N$ samples. Constructing confidence
intervals thus requires repeating this procedure many times, which can be
computationally burdensome. Nevertheless, the procedure is faster than 
previous state-of-the-art simulation-based method due to \citet{Crainiceanu.Ruppert2004}.
The latter method provides an interval for $\tau = \sigma^2_g / \sigma^2_e$,
which is straightforward to turn into an interval for $h^2 = \tau / (1 + \tau)$.

In follow-up work, \citet{schweiger2018using} proposed a stochastic
approximation to the confidence intervals of \citet{Schweigeretal2016}. The
approximation estimates the upper and lower bounds of the confidence interval
using the Robbins--Monro algorithm. Like our test-statistic, after the same
preprocessing step, the number of operations required for constructing their
intervals scales linearly with $n$.

\begin{figure}
\centering
\includegraphics[width=\linewidth]{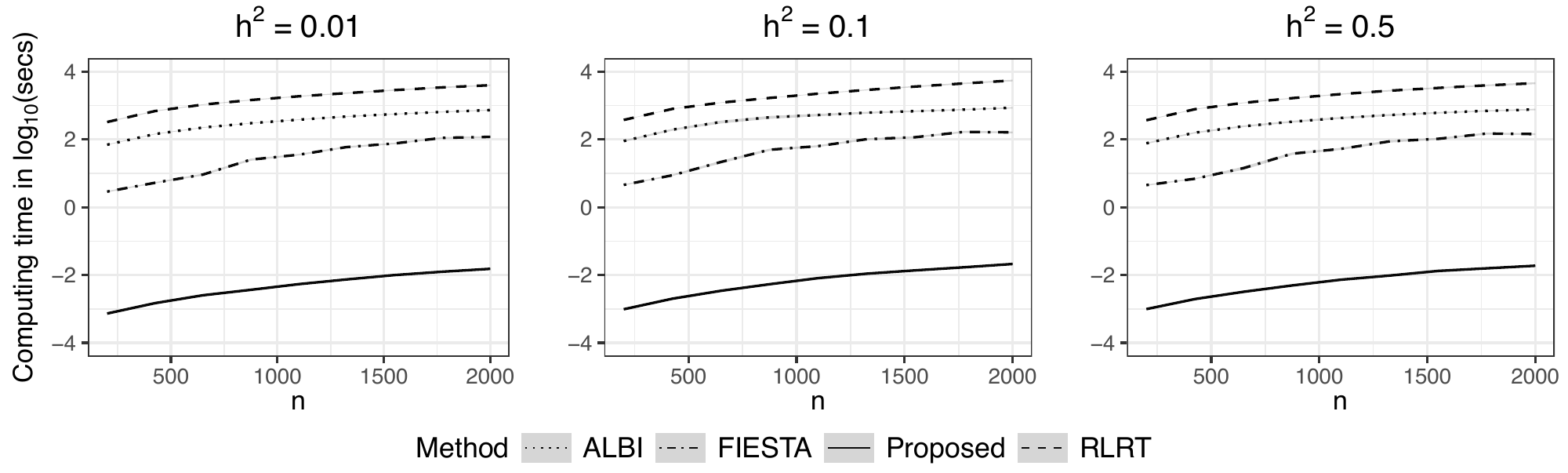}
\caption{Average computing times for one confidence interval for
      $h^2$ using the methods proposed here (``Proposed"), by
      \citet{Schweigeretal2016} (``ALBI"), by \citet{schweiger2018using}
      (``FIESTA"), and by \citet{Crainiceanu.Ruppert2004} (``RLRT''). There are shaded
      regions providing 95\% confidence bands. All times exclude the required eigendecomposition of
      $K$.  Data were generated under the same model as in
      Fig.~\ref{fig:illustration}.}
\label{fig:CompTime}
\end{figure}

In Fig.~\ref{fig:CompTime}, we compare the time needed to compute our interval
to those of \citet[ALBI,][]{Schweigeretal2016},
\citet[FIESTA,][]{schweiger2018using} and
\citet[RLRT,][]{Crainiceanu.Ruppert2004}. The figure indicates our confidence
intervals can be computed most efficiently. For example,
when $n = 200$ and $h^2 = 0.01$ our method took on average $7.3 \times 10^{-4}$
seconds compared to 2.89 seconds, 69.98 seconds and 327.10 for FIESTA, ALBI and
RLRT, respectively. When $n = 2000$ and $h^2 = 0.01$, our method took 0.02
seconds on average, whereas FIESTA, ALBI and RLRT took 119.38 seconds, 736.51
seconds and 3983.35 seconds respectively. ALBI and FIESTA are written in Python,
while our software and RLRT are written in R and C++. All timings were obtained
simultaneously on the University of Florida's high-performance computing
cluster, HiPerGator.

We also compared the widths of our confidence intervals to those of
\citet{Schweigeretal2016}.  We found that under a broad range of settings, the
distribution of the widths were similar. In the Supplementary Material, we provide plots comparing the two intervals' widths for $n \in \{200, 500, 1000, 2000\}$ and $h^2$ varying. Due to the long computing times for the interval of \citet{Crainiceanu.Ruppert2004}, we excluded it from these comparisons.

\begin{figure}
\centering
\includegraphics[width=\linewidth]{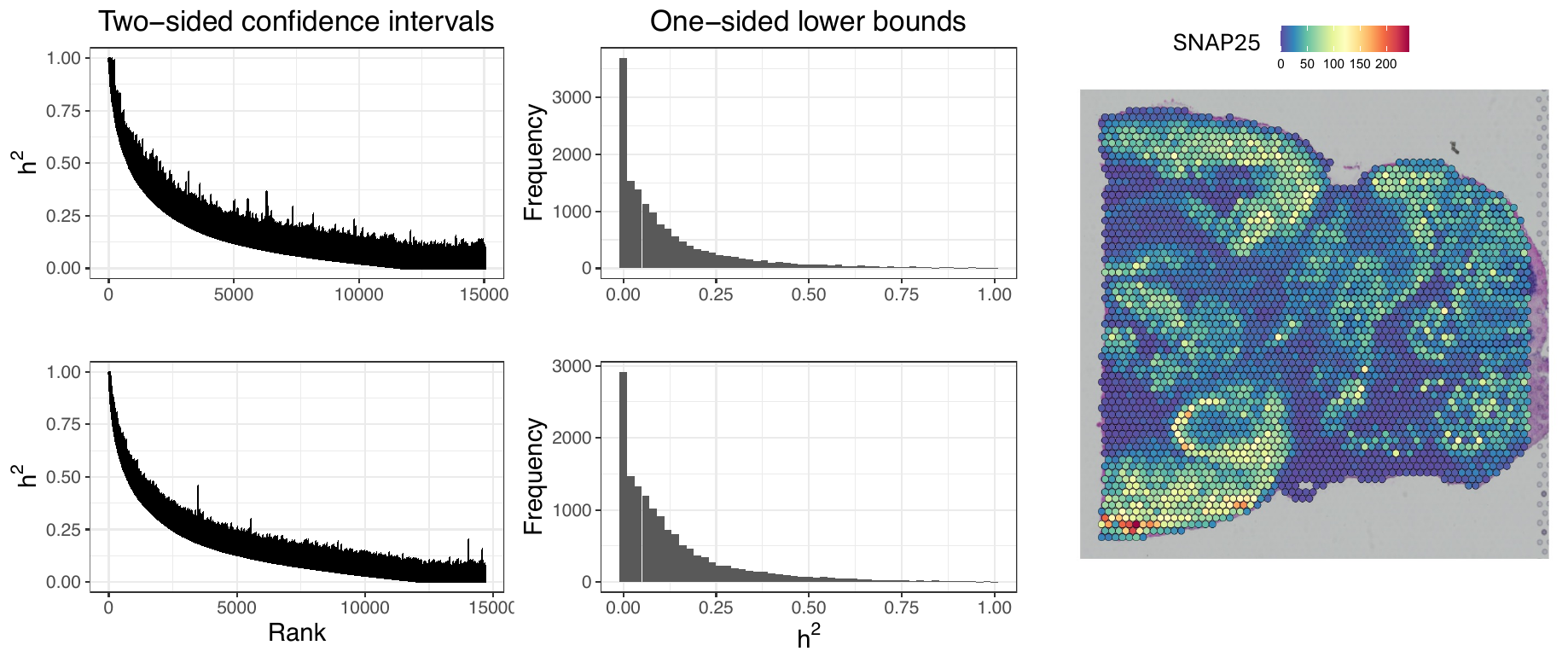}
\caption{(Left panels) Two-sided $95\%$ confidence intervals for $h^2$ in anterior (top row) and posterior (bottom row) regions of a mouse's brain tissue, sorted according to their lower bound. (Middle panels) Histograms of the lower bounds of $95\%$ one-sided confidence intervals for $h^2$ in both anterior and posterior regions. (Right panel) Expression counts of SNAP25, one of the most spatially variable genes, across the posterior region. We display counts, but our analysis is performed on log-normalized expression.}
\label{fig:ex_stx}
\end{figure}

\subsection{Gene ranking in spatially-resolved transcriptomics} \label{sec:gene_ranking}

In Section \ref{sec:Introduction}, we described one application of our method in
the analysis of spatially resolved transcriptomic data. The rightmost panels of
Fig.~\ref{fig:ex_stx} display data generated in a spatially resolved RNA
sequencing experiment performed on a mouse brain tissue sample. In such an
experiment, one measures the expression of thousands of genes (the number of RNA
reads aligning with each gene) from cells in 50 micron diameter spatial
locations (known as spots) across the tissue sample slice. These spots are
represented by the colored circles in the rightmost panels of
Fig.~\ref{fig:ex_stx}; expression is measured separately in each spot.

In this context, for one gene, $h^2$ is the fraction of variability in
expression attributed to the spatial orientation of the tissue sample. In a
typical analysis, one tests $h^2$ for more than 10,000 genes. Specifically, it
is common to assume that for the $\ell$th gene, the vector of normalized random
expression in all spots $Y_\ell$, satisfies
\begin{equation} \label{eq:geneJ}
Y_\ell \sim \mathcal{N}_n (X \beta_\ell, \sigma_\ell^2\{ h_\ell^2 K +(1-h_\ell^2) I_n\}),
\quad \ell \in \{1, \dots, d\},
\end{equation}
where $h_\ell^2$ is the fraction of spatial variability in the $\ell$th gene,
$n$ is the number of spots in the tissue, $d$ is the number of genes in the
study, and $Y_\ell$ and $Y_{\ell'}$ are independent for all $\ell \neq \ell'$.
Typically $K_{ij}= \exp(-\|s_i-s_j\|/a)$ for some fixed $a > 0$, where $s_i$ is
the two-dimensional spatial coordinates of the $i$th spot and $\|\cdot\|$ is the
Euclidean norm. 

To demonstrate the usefulness of our method in this application, we analyzed two separate
slices of a mouse's brain tissue (anterior and posterior regions), which are
available as the \textit{stxbrain} data in the \texttt{SeuratData} R package.
These data were generated on the 10X Visium platform. Assuming
\eqref{eq:geneJ} with $a= 0.02$ (chosen based on results from
\citet{weber2022nnsvg}), after quality control of filtering out low-expressed
genes and boundary spots in the tissues, $n = 2380$ for $d = 15117$ genes in the
anterior slice and $n=3011$ for $d = 14738$ genes in the posterior slice.

Using our test statistics, we can provide a confidence interval for each
$h_\ell^2$. Moreover, in order to rank genes, we can also provide one-sided
confidence intervals for $h_\ell^2$ by comparing $U_{1}(h^2, \tilde{\sigma}^2)
\{\mathcal{I}^{11}(h^2, \tilde{\sigma}^2)\}^{1/2}$ to quantiles of the standard
normal distribution. 

Because $K$ does not depend on $\ell$, we can perform a
transcriptome-wide analysis very efficiently. After a one-time
eigendecomposition of $K$, the remaining processes took around 3 minutes and 4
minutes for the two tissue regions, respectively. For comparison, ALBI took
approximately 9 minutes and 13 minutes for the two tissues, including the time
needed to find the necessary maximum likelihood estimates of each $h_\ell^2$,
which was around 3 minutes. FIESTA, which is generally slower than ALBI when
computing many confidence intervals, timed out after 96 hours.

\section*{Acknowledgement}
The authors thank an Associate Editor and two anonymous referees for comments leading to a substantially improved manuscript. The authors also thank Yu-Ru Su for a helpful conversation, and Regev Schweiger for assistance with ALBI, FIESTA, and for providing useful references. A. J. Molstad's research was supported in part by a grant from the National Science Foundation (DMS-2413294). 

\section*{Supplementary material}
\label{SM}
Proofs of lemmas and theorems are in the Supplementary Material.

\bibliographystyle{apalike}
\bibliography{main}

\end{document}